\documentclass[12pt]{article}
\usepackage{amsmath}
\usepackage{graphicx}
\usepackage{psfrag}
\usepackage{epsf}
\usepackage{enumerate}
\usepackage{natbib}
\usepackage{url} % not crucial - just used below for the URL 
\usepackage{xcolor}
\usepackage[ruled]{algorithm}
\usepackage[noend]{algpseudocode}
\usepackage{subfigure}
\usepackage{epstopdf}
\usepackage{amssymb}
\usepackage{mathtools} % for my macros
\usepackage{bm} % to bold \beta
\usepackage{bbm} % indicator function
\usepackage{comment}
\usepackage{epsfig}
\usepackage{soul}
\usepackage{booktabs}

\newcommand{\blind}{0}

\def\D{\mathbf{D}}

\begin{document}

\def\spacingset#1{\renewcommand{\baselinestretch}%
{#1}\small\normalsize} \spacingset{1}

%%%%%%%%%%%%%%%%%%%%%%%%%%%%%%%%%%%%%%%%%%%%%%%%%%%%%%%%%%%%%%%%%%%%%%%%%%%%%%

\if0\blind
{
  \title{\bf Differentially Private Methods for Releasing Results of Stability Analyses}
  \author{Chengxin Yang\thanks{The authors gratefully acknowledge the \textit{Natural Science Foundation (SES-2214756)}.} \hspace{.2cm}\\
    Department of Statistical Science, Duke University \\
    and \\
    Jerome P. Reiter\footnote{Corresponding Author: Jerome P. Reiter. Email: jreiter@duke.edu} \\
    Department of Statistical Science, Duke University}
  \maketitle
} \fi

\if1\blind
{
  \bigskip
  \bigskip
  \bigskip
  \begin{center}
    {\LARGE\bf Title}
\end{center}
  \medskip
} \fi

\bigskip
\begin{abstract}
Data stewards and analysts can promote transparent and trustworthy science and policy-making by facilitating assessments of the sensitivity of published results to alternate analysis choices.  For example, researchers may want to assess whether the results change substantially when different subsets of data points (e.g., sets formed by demographic characteristics) are used in the analysis, or when different models (e.g., with or without log transformations) are estimated on the data.  Releasing the results of such stability analyses leaks information about the data subjects. When the underlying data are confidential, the data stewards and analysts may seek to bound this information leakage.  We present methods for stability analyses that can satisfy differential privacy, a definition of data confidentiality providing such bounds. We use regression modeling as the motivating example.  The basic idea is to split the data into disjoint subsets, compute a measure summarizing the difference between the published and alternative analysis on each subset, aggregate these subset estimates, and add noise to the aggregated value to satisfy differential privacy. We illustrate the methods using regressions in which an analyst compares coefficient estimates for different groups in the data, and in which analysts fit two different models on the data. 
\end{abstract}

\noindent%
{\it Keywords:}  Confidentiality; Regression; Replicability; Reproducibility; Verification 
\vfill

\newpage
\spacingset{1.45} % DON'T change the spacing!

\section{Introduction}
\label{sec:intro}

Increasingly, researchers and policymakers demand that results of data analyses be reproducible and that scientific findings replicate beyond the original study. A minimum requirement is that, given the data and code used in a published analysis, other analysts should be able to obtain the same results as the analysts  of the original study \citep{freese2017replication}. It is also prudent to enable other analysts to assess the sensitivity of results to choices made by the original analysts, including steps in data processing and model specification \citep[e.g., see ][]{baggerley:coombes, bettis2016necessity, freese2017replication, hoeppner2019note}.
For example, the original analysts may have estimated the model using one subset of the data subjects, and it is of interest to examine whether the findings hold for other choices of subsets, e.g., across different demographic subpopulations. 
As another example, the original analysts may have made modeling choices, such as transformations (or not) or inclusion of interactions (or not), that if different would have affected the findings.  
Following the terminology in \cite{yu:kumbier}, we refer generally to these and related assessments as stability analyses.

Examples of stability analyses can be found in diverse data settings.  For example, \cite{freese2001making} examined the stability of a finding in evolutionary psychology of \cite{kanazawa2001comment} by showing how alternative regression model specifications lead to different conclusions. \cite{pollet2010no} assessed the stability of the same finding by running an analysis on alternative study data. \cite{steegen2016increasing} study  the effect of fertility on religiosity and political attitudes by examining results under different criteria for excluding study participants from the analysis. \cite{duvendack} summarize dozens of articles in economics journals that examine whether published results can be reproduced or replicated. Over 80\%  of these articles disconfirm  at least some of the originally published findings.

Effective stability analyses may require access to confidential data that cannot be disseminated with unrestricted access to the public. When so, the stewards of the confidential data must establish arrangements that provide trusted and vetted analysts access to these data; we assume these arrangements here. However, this alone may not be sufficient to protect confidentiality.  Releasing the stability analysis results themselves can introduce disclosure risks incremental to the release of original analysis results.  As a simple illustration, suppose the original analysis publishes findings that depend on the sample mean $\bar{y}$ of some sensitive variable $Y$, as well as the sample size $n$ used to compute that mean. Before computing the mean, the original analysts removed an observation $y_i$ with a particularly large value of $Y$. Seeking to check if the published findings change substantially based on the treatment of this outlier, the stability analyst computes the mean of $Y$ adding back $y_i$; call this value $\bar{y}'$. This new mean cannot be released to the public; otherwise, someone can learn $y_i$ simply by computing $y_i=(n+1)\bar{y}' - n\bar{y}$.

While the additional risk is obvious in this illustration, researchers in data privacy have shown repeatedly that releasing any statistic based on confidential data  leaks information about the data subjects' privacy. Thus, we are motivated to develop methods for stability analyses that enable data stewards and analysts to bound this information leakage.  In particular, in this article we propose stability measures that can satisfy differential privacy \citep{dwork2006calibrating, dwork2014algorithmic}. We consider two settings.  The first covers stability analyses done with alternative data (henceforth abbreviated AD), in which the stability analyst fits the same statistical model as the original analysis but with different input data, e.g., using subsets of observations from the original data or a new data source. The second covers stability analyses done with alternative models (henceforth abbreviated AM), in which the stability analyst uses the data from the original analysis but with a different model specification. We develop methods for coefficients of linear regression models, although the methods can be extended to other families of models and estimands. In doing so, we follow the framework of \cite{yu:kumbier} and develop metrics that reflect the changes in original analyses from using alternative data or alternative models.

At first glance, it may seem unnecessary to develop  differentially private algorithms for stability analysis: why not have analysts simply report differentially private results from the alternative analyses? This strategy may be reasonable in some contexts but in others not.
For some statistical analyses, existing differentially private algorithms with acceptable privacy guarantees may have high probabilities of generating large errors. \cite{barrientosdpreg} find this to be the case empirically for existing differentially private linear regression algorithms that produce estimated standard errors of the coefficients. These algorithms generally require analysts to set bounds on numerical variables that could be quite wide. For example, possible values of individual's income, a common variable for regression modeling in the social sciences, might be in the millions or billions. With large bounds, one may need very large sample sizes for the algorithms to produce accurate results \citep{barrientosdpreg}.

To assess the stability of a result, the  analyst does not necessarily need to report the estimates from the alternative analysis. It may suffice to report that the result from the alternative analysis is acceptably close (or not) to the original result. For example, rather than report $\bar{y}'$, the stability analyst could assess whether $|\bar{y} - \bar{y}'|$ is less than some practically significant difference.  We treat this as our goal, designing differentially private algorithms that offer feedback on whether the analysis result with alternative data or an alternative model is close to the result from the original analysis.

The remainder of this article is organized as follows.  In Section \ref{background}, we review differential privacy and several differentially private algorithms for perturbing outputs.  In Section \ref{methods},  we present our AD and AM methods. We describe the methods supposing the stability analysts are different  individuals than the original analysts, yet they have access to the confidential data used by the original analysts. The methods also can be used by the original analysts themselves when they include stability analyses as part of their published results, a practice consistent with recommended data science workflows \citep{yu:kumbier}. In Section \ref{workflow}, we describe some considerations for implementing these methods.  In Section \ref{illustration}, we illustrate the methods using regression analyses. Additional results from simulation studies are available in the supplemental material.  Finally, in Section \ref{conclusion}, we conclude with thoughts on future research.

\section{Background on Differential Privacy}\label{background}

Let \(\mathbf{D}\) be a data set comprising $n_0$ independently distributed individuals measured on $p$ variables, that is, \(\mathbf{D}\) is an $n_0 \times p$ matrix. Let \(\mathcal{A}\) be an algorithm (or estimator) that takes \(\mathbf{D}\) as the input and outputs a numerical quantity \(o\), so that  \(\mathcal{A}(\mathbf{D})=o\). We presume that $o$ is a discrete-valued output.  We define neighboring data sets \(\mathbf{D}'\) to differ from \(\mathbf{D}\) in only one row and identical in all other rows. In this article, we use the definition of neighboring data sets where  \(\mathbf{D}'\) and \(\mathbf{D}\) have the same number of rows.

\begin{itemize}
\item[] {\bf Definition 1 (\(\epsilon\)-differential privacy)}: 
An algorithm \(\mathcal{A}\) satisfies \(\epsilon\)-differential privacy (abbreviated as \(\epsilon\)-DP) if, for any pair of neighboring data sets \((\mathbf{D},\mathbf{D}')\) and any output \(o \in range(\mathcal{A})\), it holds that \(Pr(\mathcal{A}(\mathbf{D}) = o) \leq e^{\epsilon} Pr(\mathcal{A}(\mathbf{D}') = o)\).
\end{itemize}
 The $\epsilon$ is often called the privacy loss budget. Smaller values of \(\epsilon\) imply greater privacy guarantees.  With small $\epsilon$, differential privacy offers a strong guarantee of confidentiality protection, in that it  protects against an intruder who knows all but one observation in $\D$.

$\epsilon$-DP has appealing features that we leverage for the AD and AM stability measures \citep{dwork2014algorithmic}. 
First, $\epsilon$-DP satisfies sequential composition. When  algorithm \(\mathcal{A}_1(\mathbf{D})\) satisfies $\epsilon_1$-DP and  algorithm \(\mathcal{A}_2(\mathbf{D})\) satisfies $\epsilon_2$-DP, releasing results from the application of both on $\mathbf{D}$ satisfies $(\epsilon_1+\epsilon_2)$-DP. In other words, the privacy loss accumulates.  Thus, differential privacy offers a quantifiable bound on the amount of information leaked from releasing results of multiple queries on $\D$.  Second, $\epsilon$-DP satisfies parallel composition.  For data files $\mathbf{D}_1$ and  $\mathbf{D}_2$ with disjoint sets of individuals, and algorithms \(\mathcal{A}_1(\mathbf{D})\) and \(\mathcal{A}_2(\mathbf{D})\) satisfying $\epsilon_1$-DP and  $\epsilon_2$-DP, respectively, releasing the results of \(\mathcal{A}_1(\mathbf{D})\) and \(\mathcal{A}_2(\mathbf{D})\) satisfies $\min(\epsilon_1, \epsilon_2)$-DP.  Parallel composition implies that we can apply differential privacy on disjoint data sets without compounding the information leakage.  Finally, $\epsilon$-DP satisfies parallel composition.  Given an algorithm \(\mathcal{A}\) that satisfies $\epsilon$-DP and any algorithm \(\mathcal{A}^*\) that takes $\mathcal{A}(\mathbf{D})$ as the input, releasing \(\mathcal{A}^*(\mathcal{A}(\mathbf{D}))\) also satisfies \(\epsilon\)-DP.  In other words, manipulating a differntially private output does not affect the privacy guarantee.
We use the post-processing property to facilitate  interpretations of our stability measures.

One method to ensure \(\epsilon\)-DP is the Laplace Mechanism \citep{dwork2006calibrating}. Let $f$ be some function that we apply on $\mathbf{D}$, such as a sum of elements or the least squares estimates of regression coefficients.  Absent privacy concerns, we would compute and release $f(\mathbf{D})$. Define the global sensitivity of \(f\) to be \(\Delta(f) = max_{(\mathbf{D},\mathbf{D}')}||f(\mathbf{D})-f(\mathbf{D}')||\) where \((\mathbf{D},\mathbf{D}')\) is any pair of neighboring data sets. Thus, $\Delta(f)$ is the maximum amount $f(\mathbf{D})$ can change if we change one observation in any $\mathbf{D}$.  For example, if $f$ is the sum function and $\mathbf{D}$ includes a single binary variable, then $\Delta(f) =1$.  On the other hand, if $f$ computes the mean of a non-negative integer-valued variable that can take a maximum value of $B$, then $\Delta(f) = B/n$, where $n$ is the number of terms in the mean. The global sensitivity is used in the Laplace Mechanism, defined as follows.

\begin{itemize}
\item[]  {\bf Definition 2 (Laplace Mechanism)}: The Laplace Mechanism releases \(f(\mathbf{D}) + \eta\), where \(\eta \sim Laplace(0,\Delta(f)/\epsilon)\).  It satisfies \(\epsilon\)-DP.
\end{itemize}

For some functions, the $\Delta(f)$ can be sizeable.  A large global sensitivity can correspond to a large variance in the Laplace distribution, resulting in high probability of generating an unacceptably large $\eta$. For example, suppose feasible values for individuals' incomes are in $[0, 1000000000]$, and that $n=10000$.  Then, $\Delta(f)= 10000$  for the sample mean income. Assuming $\epsilon=1$, a draw $\eta \sim Laplace(0,10000)$ has  almost  60\% probability of exceeding 5000 in absolute value and more than  20\% probability of exceeding 16000 in absolute value. The average income in the United States is around \$70000, with differences by subpopulations, e.g.,  defined by sex, age, or state, typically on the order of \$5000 to \$16000 \citep{forbes}.  Thus, given a set of noisy  versions of subpopulation sample means produced from this Laplace Mechanism, analysts likely cannot tell whether differences (large or small) in the noisy means reflect actual disparities (or actual closeness) or simply distortions from the privacy protection. This exemplifies why analysts may find direct application of $\epsilon$-DP to estimates not suitable for stability analysis.

As this illustration suggests, when $\Delta(f)$ is sizeable, it  may be preferable to use alternative mechanisms \citep{Hawes2020Implementing}. 
One approach is the sub-sample and aggregate algorithm \citep{nissim2007smooth}. The idea is to reduce the sensitivity of \(f\) by partitioning \(\mathbf{D}\) into \(M\) disjoint subsets, \(\{\mathbf{D}_{l}:l=1, \dots, M\}\).
We  evaluate \(f(\mathbf{D}_{l})\) on each \(\mathbf{D}_{l}\), and compute their average $f^{new}(\mathbf{D}) = \sum_{l=1}^M f(\mathbf{D}_{l})/M$. For many $f$, changing one observation changes at most one of the partitions and the corresponding $f(\mathbf{D}_{l})$; thus, $\Delta(f^{new}) = \Delta(f)/M$. We can release \(f^{new}(\mathbf{D})+ \eta^{new}\), where \(\eta^{new} \sim Laplace(0,\Delta(f)/M\epsilon)\). This can reduce the variance of the added noise substantially, making it less likely to generate noise that obscures $f^{new}(\mathbf{D})$. A downside, however, is that one must interpret  $f^{new}(\mathbf{D})$ rather than  $f(\mathbf{D})$, and  it is not necessarily the case that $f^{new}(\mathbf{D}) = f(\mathbf{D})$. We use the sub-sample and aggregate algorithm to construct differentially private stability measures, as we now describe.

\section{Differentially Private Algorithms for Stability Analysis}\label{methods}

We work in the setting of linear regressions. Suppose  the \(\mathbf{D}\) used in the original data analysis comprises \(n_0\) individuals measured on \(p\) explanatory variables \(X_1, \dots, X_{p}\) and a response \(Y\). The original researchers estimated some linear regression model, which we refer to as \(Model_0\), using \(\mathbf{D}\) and published its estimated coefficients and standard errors.  Among these coefficients is a particular variable of scientific interest \(X \in \{X_1, \dots, X_p\}\); for example, $X$ could be an indicator variable representing the presence or absence of some treatment or intervention.  Let  \(\hat{\gamma}_o\) be the published estimate of \(\gamma_o\), which we define as the true coefficient of \(X\) in \(Model_0\) in the population represented by $\mathbf{D}$. For now, we do not assume any specific privacy policies on the release of the output from \(Model_0\) estimated with $\mathbf{D}$, including \(\hat{\gamma}_o\).  We discuss the implications of the privacy policy for \(\hat{\gamma}_o\), e.g., whether the released statistic is differentially private or not, in Section \ref{sec:privguar}.

Let $\mathbf{D}^*$ be the data set used for stability analyses, comprising $N$ individuals. The  \(\mathbf{D}^*\) could be the same as or a proper subset of  \(\mathbf{D}\), i.e., $\mathbf{D}^* \subseteq \mathbf{D}$, with an example of the latter being the data for individuals in $\mathbf{D}$ who belong to a certain subpopulation. It also could be a data set that has the variables used in \(Model_0\) but no individuals from \(\mathbf{D}\), such as a sample taken from a different location.  Regardless, we assume that record-level information in $\mathbf{D}^*$ is confidential, and that the data steward requires any new statistics computed with $\mathbf{D}^*$ to come with a privacy guarantee. We assume the stability analyst has access to all records in $\mathbf{D}^*$ as well as those in $\mathbf{D}$, for example, via a secure data enclave. We discuss AD measures for settings where we add individuals to $\mathbf{D}$, i.e., $\mathbf{D} \subset \mathbf{D}^*$, in Section \ref{conclusion}.

We now describe the AD and AM stability measures. We first  describe the methodologies in Section \ref{sec:AD} and Section \ref{sec:AM},  deferring a discussion of the  privacy guarantees until Section \ref{sec:privguar}.
For convenient reference, we summarize notation for the AD measures in Table \ref{tab:AD:notations} and for the AM measures in Table \ref{tab:AM:notations}.

\subsection{Methods for AD analyses}\label{sec:AD}

\begin{table}[t]
\centering
\begin{tabular}{cl} \toprule
Notation & Explanation \\ \midrule
$\mathbf{D}$ & Data for original analysis \\
$\mathbf{D}^*$ & Data for stability analysis \\
$\mathbf{D}^*_l$ & The $l$-th partition of $\mathbf{D}^*$, where $l \in \{1, \dots, M\}$ \\
$n_0$ & Sample size of $\mathbf{D}$ \\
$N$ & Sample size of $\mathbf{D}^*$ \\
$n$ & Sample size(s) of $\mathbf{D}^*_l$, allowing one more or less for some partitions \\
$\gamma_o$ & True coefficient of interest in population represented by $\mathbf{D}$ \\ 
$\gamma_r$ & True coefficient of interest in population represented by $\mathbf{D}^*$ \\
$\hat{\gamma}_o$ & Estimate of $\gamma_o$ computed using $\mathbf{D}$ \\
$\hat{\gamma}_r$ & Estimate of $\gamma_r$ computed using $\mathbf{D}^*$ \\
$\hat{\gamma}_{r,l}$ & Estimate of $\gamma_r$ computed using $\mathbf{D}^*_l$ \\
$T(\hat{\gamma}_o;\boldsymbol{\alpha}_T)$ & Tolerance region \\
$U(\hat{\gamma}_o;\boldsymbol{\alpha})$ & Adjusted tolerance region used with each   $\mathbf{D}^*_l$ \\
$S$ & Number of $\hat{\gamma}_{r,l} \in U(\hat{\gamma}_o;\boldsymbol{\alpha})$ for all $l$ \\
$S^R$ & $\epsilon$-DP version of $S$. $S^R = (S + \eta)$ and $\eta \sim Laplace(0,1/(M\epsilon))$ \\
$r$ & Probability of $\hat{\gamma}_{r,l} \in U(\hat{\gamma}_o;\boldsymbol{\alpha})$ \\
\bottomrule
\end{tabular}
\caption{Summary of notation for the AD stability measures.} \label{tab:AD:notations}
\end{table}

Let $\gamma_r$ be the true coefficient of $X$ when fitting  \(Model_0\) on the population defined by \(\mathbf{D}^*\), and let \(\hat{\gamma}_r\) be the estimate of  \(\gamma_r\) computed with $\mathbf{D}^*$.  Regardless of the nature of $\mathbf{D}^*$, for AD analyses it is salient to consider differences between \(\hat{\gamma}_r\) and \(\hat{\gamma}_o\).  When $\mathbf{D}^*$ comprises a disjoint set of individuals, large differences suggest that \(\hat{\gamma}_o\) does not accurately describe the relationship between $X$ and $Y$ beyond the original study. We call this a stability failure. When \(\mathbf{D}^* \subset \mathbf{D}\), large differences suggest that the original estimate \(\hat{\gamma}_o\) is overly sensitive to the choice of records used to estimate the model, which we consider another type of stability failure.

More specifically, we conceive of an analyst who has in mind some tolerance region  \(T(\hat{\gamma}_o;\boldsymbol{\alpha}_T)\) for assessing the stability of the original analysis, where \(\boldsymbol{\alpha}_T\) are any parameters defining the region. When \(\hat{\gamma}_r \in T(\hat{\gamma}_o;\boldsymbol{\alpha}_T)\), the analyst concludes that the original result is stable; otherwise, the analyst concludes it is not stable. Naturally, the specification of \(T(\hat{\gamma}_o;\boldsymbol{\alpha}_T)\) significantly influences this decision. We discuss specification of tolerance regions in Section \ref{sec:workflowAD}.

Under differential privacy, we cannot directly release the indicator of whether \(\hat{\gamma}_r \in T(\hat{\gamma}_o;\boldsymbol{\alpha}_T)\).  However, stability analysts should not simply apply a Laplace Mechanism to the indicator. To see this, let $f(\mathbf{D}^*)$ be this indicator. The $\Delta(f)=1$, since modifying one observation in $\mathbf{D}^*$ could move $\hat{\gamma}_r$ in or out of $T(\hat{\gamma}_o;\boldsymbol{\alpha}_T).$  As a result, for small $\epsilon$, the amount of noise added likely would obscure whether the indicator is one or zero, defeating its purpose. For example, with $\epsilon =1$, there is a 60\% chance of drawing $|\eta| > 0.5$; with $\epsilon=0.5$ this probability rises to nearly 80\%.   Instead, following the strategy used in \cite{barrientos2018providing} and \cite{Amitai_Reiter_2018}, we use the sub-sample and aggregate technique to define measures over partitions of $\mathbf{D}^*$ rather than the entirety of $\mathbf{D}^*$ itself.

The stability analyst randomly partitions \(\mathbf{D}^*\) into \(M\) disjoint subsets, \(\mathcal{P}=\{\mathbf{D}^*_1, \dots, \mathbf{D}^*_M\}\) of size $n=\lfloor{N/M}\rfloor$, allowing some partitions to have one more or fewer observations as necessary; we discuss the choice of $M$ in Section \ref{sec:workflowAD}.
In each \(\mathbf{D}^*_l\), the analyst computes the estimate of \(\gamma_r\), which we denote \(\hat{\gamma}_{r,l}\). The analyst specifies a tolerance region $U(\hat{\gamma}_o;\boldsymbol{\alpha})$ for use with $\mathcal{P}.$ For some stability analyses, the  analyst may set $U(\hat{\gamma}_o;\boldsymbol{\alpha})=T(\hat{\gamma}_o;\boldsymbol{\alpha}_T)$. In other contexts, the analyst may adjust $T(\hat{\gamma}_o;\boldsymbol{\alpha}_T)$ to account for the smaller sample size, as discussed in Section \ref{sec:workflowAD}. For each \(\mathbf{D}^*_l\), let \(W_l = \mathbb{I}(\hat{\gamma}_{r,l} \in U(\hat{\gamma}_o;\boldsymbol{\alpha}))\), where \(\mathbb{I}(A)=1\) when \(A\) is true and \(\mathbb{I}(A)=0\) otherwise. Values of $W_l=1$ favor stability and $W_l=0$ do not.  Essentially, the analyst has the outcomes $\mathbf{W}=(W_1, \dots, W_M)$ of $M$ stability studies on disjoint data sets of size $n$. We summarize $\mathbf{W}$ using \(S = \sum_{l=1}^M W_l\). Thus, values of $S$ near $M$ favor stability and values near zero do not.
 
We cannot release $S$ (or $\mathbf{W}$) as is if we wish to satisfy $\epsilon$-DP.  We therefore apply the Laplace Mechanism on \(S\), resulting in \(S^R = S + \eta\) where \(\eta \sim Laplace(0,1/\epsilon)\). The sensitivity of \(S\) equals 1 because, considering $\hat{\gamma}_o$ as constant, changing one observation in \(\mathbf{D}^*\) can switch at most one $W_l$ from 1 to 0 or vice versa. Although the analyst could publish $S^R/M$ as the output of the stability analysis, we take an additional, post-processing step---which has no effect on the privacy guarantee---to enhance interpretability. We model each \(W_l\) as an independent draw from a \(Bernoulli(r)\), where $r$ is the probability that $\hat{\gamma}_{r,l} \in U(\hat{\gamma}_o;\boldsymbol{\alpha})$. We then treat $r$ and $S$ as unobserved random variables, using the perturbed value of $S^R$ as the observed data, and compute the Bayesian posterior distribution of $r$.  Specifically, we let \[S^R | S \sim Laplace(S, 1/\epsilon), \ S|r \sim Binomial(M,r), \ r \sim \phi_0 ,\] where \(\phi_0\) is a prior distribution for \(r\).  Here, we use a uniform prior distribution.  We compute the posterior distribution of $r$ using Markov chain Monte Carlo (MCMC) sampling. The stability analyst can report summaries of the posterior distribution, such as the posterior mode, credible intervals, or the $P(r > \delta \mid S^R)$ for some user-specified \(\delta\), e.g., \(\delta = 1/2\). The analyst also should summarize $\mathbf{D}^*$ in broad terms, e.g., name the subgroup being considered, and the limits of the tolerance interval.

It is important to note that $r$ is not the $Pr(\hat{\gamma}_r \in T(\hat{\gamma}_o;\boldsymbol{\alpha}_T))$. However, when $U(\hat{\gamma}_o;\boldsymbol{\alpha})$ is large compared to the sampling variance due to partitioning in $\hat{\gamma}_{r,l},$ for fixed AD measures it is reasonable to interpret $r$ as an estimate of $Pr(\hat{\gamma}_r \in T(\hat{\gamma}_o;\boldsymbol{\alpha}_T))$. We discuss  simulation-based approaches to facilitate the interpretations of $r$ in Section \ref{sec:workflowAD}.

\subsection{Methods for AM analyses}\label{sec:AM}

\begin{table}[t]
\centering
\begin{tabular}{cl} \toprule
Notation & Explanation \\ \midrule
$\mathbf{D}$, $\mathbf{D}^*$ & Data for original and stability analyses \\
$\mathbf{D}_l$ & The $l$-th partition of $\mathbf{D}$, where $l \in \{1, \dots, M\}$ \\
$\beta$ & True coefficient of interest in $Model_0$ (original analyses) \\ 
$\gamma$ & True coefficient of interest in $Model_1$ (stability analyses) \\
$CI_{\beta}$, $CI_{\gamma}$ & $\alpha$-level confidence intervals for $\beta$ and $\gamma$ using $\mathbf{D}$ \\
$CI_{\beta,l}$, $CI_{\gamma,l}$ & $\alpha$-level confidence intervals for $\beta$ and $\gamma$ using $\mathbf{D}_l$, where $l \in \{1, \dots, M\}$ \\
$\nu$ & Overlap measure between $CI_{\beta}$ and $CI_{\gamma}$ \\
$\bar{\nu}$ & Average of the overlap measure between $CI_{\beta,l}$ and $CI_{\gamma,l}$ over all $l$ \\
$\bar{\nu}^R$ & $\epsilon$-DP version of $\bar{\nu}$. $\bar{\nu}^R = \bar{\nu} + \eta_{\nu}$ and $\eta_{\nu} \sim Laplace(0,1/(M\epsilon))$ \\
\bottomrule
\end{tabular}
\caption{Summary of notation for the AM stability measures.} \label{tab:AM:notations}
\end{table}

For AM stability assessments, the analyst seeks to assess how conclusions about the coefficient of the effect of the variable of interest $X$ change when using an alternative model, say \(Model_1\), versus using \(Model_0\).  We assume the analyst estimates both models on the original data set, i.e.,  \(\mathbf{D} = \mathbf{D}^*\).
In addition, we assume \(Model_1\) includes $X$ in the linear predictor with the same functional form as \(Model_0\), e.g., $X$ is an indicator variable in both models.  \(Model_1\) has a different functional form for other predictors, for example, including (excluding) some variable in $\mathbf{D}^*$ that was not (was) in \(Model_0\), or using transformations of some predictors.   Let \(\beta\) and $\gamma$ be the coefficients of \(X\)  in \(Model_1\) and \(Model_0\), respectively, for the population represented by \(\mathbf{D}\).  Since since \(Model_1\) uses a different specification, \(\beta\) and $\gamma$ are not necessarily the same quantity.  For example, the slope of $X$ in the regression of $Y$ on $(X, X_1)$ differs from the slope of $X$ in the regression of $Y$ on $(X, X_1, X_2)$, unless $X$ is independent of $(X_1, X_2)$.
Thus, for AM analyses, our goal is to compare $\beta$ and $\gamma$.

We do so by comparing not just their point estimates but their interval estimates, so as to account for sampling uncertainty when assessing the stability of results to the different model specifications  \citep{hoeppner2019note}. In particular, we assess the overlap in confidence intervals (CIs) for the two coefficients \citep{karr2006framework, wheeler2006comparing, maghsoodloo2010comparing,knol2011mis}.  When the CIs overlap substantially, we infer that using \(Model_1\) or \(Model_0\) leads to similar conclusions about the relationship between $X$ and $Y$.  When the CIs do not overlap substantially, we infer that \(Model_1\) leads to different conclusions about that relationship. Of course, even with non-overlapping CIs, the differences in $\beta$ and $\gamma$ may not be practically significant. In such cases, the stability analyst may wish to supplement the AM analysis with a differentially private comparison of the point estimates in the two models; we suggest a method for doing so after defining the AM method below. 

To measure CI overlap, we use the metric proposed by \cite{karr2006framework} in the context of statistical disclosure limitation; see \eqref{eq:CIoverlap} below. This metric takes a maximum value of one when the CIs overlap completely and a  minimum value of zero when the two intervals do not overlap at all.  

To satisfy $\epsilon$-DP,  we cannot release the CI overlap directly, as it is a deterministic function of the values in $\mathbf{D}^*$. However, we should not use the Laplace Mechanism to perturb its value. The global sensitivity of the overlap measure equals 1, as it is hypothetically possible for the modification of one highly influential point to change the intervals from perfectly overlapping to completely not.  Therefore, adding Laplace noise easily could obscure the information in the CI overlap measure.  Instead, we again turn to the sub-sample and aggregate algorithm to create a differentially private CI overlap measure.

We randomly partition $\mathbf{D}^*$ as in Section \ref{sec:AD}.  In each \(\mathbf{D}^*_l\), we fit both  \(Model_0\) and \(Model_1\), and compute the corresponding \(\alpha\)-level CIs for the coefficients of $X$.  Henceforth, we assume \(\alpha=95\%\). For each $\mathbf{D}^*_l$, let $CI_{\gamma,l}$ be the CI for $\gamma$  with lower limit $L_{\gamma, l}$ and upper limit  $U_{\gamma,l}$, so that \(CI_{\gamma,l} = (L_{\gamma, l}, U_{\gamma,l})\); and, let $CI_{\beta,l}$ be the CI for $\beta$  with lower limit $L_{\beta, l}$ and upper limit  $U_{\beta,l}$, so that \(CI_{\beta,l} = (L_{\beta, l}, U_{\beta,l})\). When \(CI_{\gamma,l} \cap CI_{\beta,l} \neq \emptyset\), define \([L_l, U_l]\) as \(CI_{\gamma,l} \cap CI_{\beta,l}\), otherwise define \((L_l, U_l)\) as \((0,0)\). Then, we compute the overlap between the two intervals, 
\begin{equation}\nu_l = \frac{1}{2} \left(\frac{U_l-L_l}{U_{\gamma,l}-L_{\gamma,l}}+\frac{U_l-L_l}{U_{\beta,l}-L_{\beta,l}}\right).\label{eq:CIoverlap}
\end{equation}
Values of $\nu_l$ near one indicate that the two models produce similar interval estimates for the coefficient of $X$ on data sets of size $n$, which supports stability.  We summarize $(\nu_1, \dots, \nu_M)$ with their average, \(\bar{\nu} = \sum_{l=1}^M \nu_{l}/M\), which represents the amount of CI overlap expected in samples of size $n$ when using \(Model_1\) and \(Model_0\) to estimate the coefficient of $X$.

In keeping with $\epsilon$-DP, we should not release $\bar{\nu}$.  We instead use a Laplace Mechanism to create $\nu^R = \bar{\nu} + \eta_{\nu}$.  Since the global sensitivity of $\bar{\nu}$ is \(1/M\), we can sample $\eta_{\nu} \sim Laplace(0, 1/\epsilon M)$.  To further aid interpretation, we treat $\bar{\nu}$ as unknown and compute its posterior distribution using 
\[\nu^R | \bar{\nu} \sim Laplace(\bar{\nu}, 1/M\epsilon), \ \bar{\nu} \sim \psi_0,\]
where \(\psi_0\) is the prior distribution for \(\bar{\nu}\).  With a Beta($a,b$) prior distribution,  we can sample directly from the posterior distribution,
\begin{equation}\label{eq:nubar}
p(\bar{\nu}|\bar{\nu}^R) \propto e^{-M\epsilon |\bar{\nu}-\bar{\nu}^R|} \bar{\nu}^{a-1} (1-\bar{\nu})^{b-1} \mathbb{I}(0 \leq \bar{\nu} \leq 1).
\end{equation}
The analyst can report summaries of this posterior distribution, e.g., its mode and 95\% credible intervals, as the outputs of a stability analysis. The analyst also should report the specifications (without parameter estimates) in \(Model_1\) and \(Model_0\) as part of the output.

When the CIs do not overlap, the analyst may want to further assess and report on the differences in the estimates of $\gamma$ and $\beta$. Let $\hat{\gamma}_{r,l}$ and $\hat{\beta}_{r,l}$ be the point estimates of $\gamma$ and $\beta$ computed in $\mathbf{D}^*_l$, where $l=1, \dots, M$.  To assess differences in the point estimates, the analyst can replace \eqref{eq:CIoverlap} with a tolerance region for differences  within partitions, for example, $|\hat{\gamma}_{r,l} - \hat{\beta}_{r,l}| < K$. Following the logic of the AD methods, the analyst can set $W_l=1$ when the difference satisfies the tolerance region and set $W_l=0$ otherwise. Using the resulting $(W_1, \dots, W_M)$, the analyst can follow the strategy used in the AD methods to compute $S^R$ and the posterior distribution of $r$.

\subsection{Privacy protection properties}
\label{sec:privguar}

The use of differentially private algorithms in Section \ref{sec:AD} and Section \ref{sec:AM}  is intended to manage the incremental disclosure risks inherent in releasing additional results from stability analyses.  However, the precise nature of the privacy protection depends on how results from the original analysis of $\mathbf{D}$ were released, as well as the relationship between $\mathbf{D}$ and $\mathbf{D}^*$,  as we now describe. Here, we suppose that the stability analysis is released using $\epsilon$ as the privacy budget, the stability analyst has access to $\mathbf{D}$ and $\mathbf{D}^*$, and a set of original results from $\mathbf{D}$ has been published already.

We first consider the case where results from the original analysis of $\mathbf{D}$ do not satisfy $\epsilon$-DP.  This is a common scenario; for example, currently outputs derived from analyses inside the Federal Statistical Research Data Centers are released after disclosure control treatment that does not satisfy $\epsilon$-DP.  When $\mathbf{D} = \mathbf{D}^*$ as in the AM setting, the release of the stability measure itself is differentially private. However, the overall privacy loss due to the release of the results from $\mathbf{D}$ and the AM stability measure is difficult to quantify, since we generally cannot take advantage of the sequential composition property of $\epsilon$-DP when the release of the original analysis of $\mathbf{D}$ is not differentially private. When $\mathbf{D}^*$ is a data set of individuals who are not in  $\mathbf{D}$, as in the AD stability analysis with a new data set, the released stability analysis satisfies $\epsilon$-DP for the records in $\mathbf{D}^*$.  It does not release any additional information about the records in $\mathbf{D}$ beyond what has been released previously.  When $\mathbf{D}^* \subset \mathbf{D}$,  as in the AD analysis of a subpopulation, the  privacy guarantee is on the incremental disclosure risk to individuals in $\mathbf{D}^*$ from releasing outputs of stability analyses, given whatever is released in the original analysis.  However, we do not have sequential composition; thus, we cannot easily ascertain the overall disclosure risk from releasing both sets of outputs. Nonetheless, in this case, we expect the disclosure risks from releasing the differentially private stability measures  generally to be reduced compared to releasing stability measures without any privacy protection.

We next consider the case where the release of results from the original analysis of $\mathbf{D}$ satisfies $\epsilon_1$-DP. When $\mathbf{D}=\mathbf{D}^*$, the sequential composition property of $\epsilon$-DP guarantees that the released AM stability analysis will satisfy $(\epsilon+\epsilon_1)$-DP.  For the AD setting, we  presume that the stability analyst uses the differentially private output from the original analysis of $\mathbf{D}$ when computing the AD stability measures.  With this presumption, when $\mathbf{D}^*$ is a data set of individuals that is disjoint with  $\mathbf{D}$,  the parallel composition property of differential privacy assures that the combined release satisfies $\min(\epsilon, \epsilon_1)$-DP. When $\mathbf{D}^* \subset \mathbf{D}$, the additional release of the AD stability measure  will satisfy $(\epsilon+\epsilon_1)$-DP.

It is worth noting that, in practice, stability analyses are likely to involve exploratory analyses that the analyst does not intend to publish.  For example, the analyst may examine multiple subpopulations or multiple model specifications, yet only seek to publish stability analyses from a selection of these exploratory analyses.  This technically violates the $\epsilon$-DP guarantee, unless all of these exploratory analyses are themselves differentially private \citep{dworkfienberg}.  Allowing for exploratory data analysis with formal privacy guarantees is an open question for future research.

\section{Implementing the AD and AM  Measures}\label{workflow}

In this section, we offer guidance on several issues related to implementing the AD and AM measures of Section \ref{methods}.  We do not address the choice of \(\epsilon\), as we assume it  is mandated by the holders of $\mathbf{D}^*$.

\subsection{Implementation of AD measures}\label{sec:workflowAD}

Three key decisions in AD stability assessments include the specification of the tolerance region, the number of partitions $M$, and, when summarizing the posterior distribution of $r$, the value \(\delta\) that represents evidence of stability.  We now discuss considerations for each of these specifications.

\subsubsection{Specifying the tolerance region}

The analyst should specify a tolerance region that reflects their desiderata for claiming stability.  For example, the analyst may decide that results are adequately stable if $\hat{\gamma}_r$ and $\hat{\gamma}_o$ have the same sign. If so, the analyst could set $T(\hat{\gamma}_o;\boldsymbol{\alpha}_T) = %U(\hat{\gamma}_o;\boldsymbol{\alpha}) = 
\textrm{sign}(\hat{\gamma}_o)(0, \infty)$. 
 For the differentially private version of this stability measure, the analyst also could set 
$U(\hat{\gamma}_o;\boldsymbol{\alpha})= \textrm{sign}(\hat{\gamma}_o)(0, \infty)$.
Another example is given by $T(\hat{\gamma}_o;\boldsymbol{\alpha}_T) = U(\hat{\gamma}_o; \boldsymbol{\alpha}) = [\hat{\gamma}_o \pm \alpha|\hat{\gamma}_o|]$, operationalizing whether $\hat{\gamma}_r$ is within $\alpha$ percentage of $\hat{\gamma}_o$. More generally, the analyst could set $T(\hat{\gamma}_o;\boldsymbol{\alpha}_T) =U(\hat{\gamma}_o;\boldsymbol{\alpha})= (c, d)$, where $c$ and $d$ are plausible values of $\hat{\gamma}_r$.  This type of region---also used by \citet{barrientos2018providing} to evaluate the quality of synthetic data \citep{raghunathan2003multiple, mcclure2012towards}---provides evidence of the location of $\hat{\gamma}_r$. 
We refer to tolerance regions where $T(\hat{\gamma}_o;\boldsymbol{\alpha}_T) = U(\hat{\gamma}_o;\boldsymbol{\alpha})$ as fixed tolerance regions, in that they do not change by basing stability checks on $\mathbf{D}^*$ or $\mathcal{P}$.

As an alternative, analysts could use what we call an adjusted tolerance region, where $U(\hat{\gamma}_o;\boldsymbol{\alpha})$ is a modification of $T(\hat{\gamma}_o;\boldsymbol{\alpha}_T)$.  For example, analysts could decide that results are stable when $\hat{\gamma}_r$ is within a certain distance of $\hat{\gamma}_o$, allowing for larger distance when using $\mathcal{P}$, which is based on a sample of size $n$, than they would using $\mathcal{D}^*$, which is based on a sample of size $N$. To illustrate, suppose the stability analyst has in mind a tolerance region based on the published standard error of $\hat{\gamma}_o$, which we write as $\hat{\sigma}(\hat{\gamma}_o)$; for example, 
\begin{equation}\label{eq:T:AD:CI}
T(\hat{\gamma}_o;\alpha_T) = [\hat{\gamma}_o - \alpha_T \hat{\sigma}(\hat{\gamma}_o), \hat{\gamma}_o + \alpha_T \hat{\sigma}(\hat{\gamma}_o)].    
\end{equation}
Here, \(\alpha_T\) determines the number of standard errors difference that is tolerated for stability. When using $\mathcal{P}$, $\hat{\gamma}_{r,l}$ typically has larger standard error than $\hat{\gamma}_r$.  The increased uncertainty could result in high probability that any \(W_l=0\), even if $\hat{\gamma}_r$ is very close to $\hat{\gamma}_o$, potentially leading the analyst to a dubious conclusion.  Thus, the analyst may prefer to adjust $T(\hat{\gamma}_o;\alpha_T)$.

For AM measures based on \eqref{eq:T:AD:CI}, we suggest replacing $\hat{\sigma}(\hat{\gamma}_o)$ with an approximate standard error for $\hat{\gamma}_{r,l}$, which we write as $\hat{\sigma}(\hat{\gamma}_r) =  \sqrt{n_0/n}\hat{\sigma}(\hat{\gamma}_o)$, where $n_0$ is the sample size of  $\mathbf{D}$ used to compute $\hat{\gamma}_o$. This type of variance adjustment was proposed in \cite{barrientos2019differentially} in the context of significance testing for regression coefficients.  Thus, the stability analyst can use the adjusted tolerance region,  \begin{equation}\label{eq:U:AD:CI}
U(\hat{\gamma}_o;\alpha) = [\hat{\gamma}_o - \alpha\sqrt{n_0/n}\hat{\sigma}(\hat{\gamma}_o), \hat{\gamma}_o + \alpha\sqrt{n_0/n}\hat{\sigma}(\hat{\gamma}_o)].
\end{equation}

\subsubsection{Choosing $M$}\label{sec:AD:M}

In addition to selecting the tolerance region, the stability analyst needs to select $M$ and, consequently, $n$.  Obviously, for a fixed $\mathbf{D}^*$, choosing a larger $M$  results in a smaller $n$. Generally, using a larger $M$ reduces the variance of \(S/M\) and hence the size of the error from the Laplace Mechanism relative to $S/M$, while simultaneously increasing the variance of each \(\hat{\gamma}_{r,l}\) because of the smaller $n$.  These two effects typically pull in opposite directions.  With small $n$, the variance of \(\hat{\gamma}_{r,l}\) could be so large that $S/M$ itself is not informative as a stability measure. On the other hand, with small \(M\), the contribution to the variance of $S^R/M$ from  the Laplace Mechanism can be high enough to obscure $S/M$ and hence the conclusions from the stability analyses. These effects are illustrated  in the simulations in the supplementary material.

When selecting $M$ and $n$, we recommend that analysts consider the following points.  First, analysts should ensure that the sample space of \(S/M\) is detailed enough to facilitate interpretations.  For example, when \(M = 5\), \(S/M\) takes values in \(\{0, .2, .4, .6, .8, 1\}\), which may offer too crude an interpretation for the stability assessment at hand. For most contexts, we expect that analysts would desire gradations of at least .05, i.e.,  set \(M \geq 20\). Second, as best as possible analysts should prevent the Laplace Mechanism from injecting large error with high probability, which further supports not letting $M$ be too small.

Beyond these considerations, a simulation-based approach can help with choosing $M$.  Ideally, the sampling distribution of $S^R/M$ has most of its mass near zero---or below some threshold $\delta>0$ chosen by the analyst---when $\hat{\gamma}_r$ is far from the boundary points and outside of $T(\hat{\gamma}_o;\boldsymbol{\alpha}_T)$, and it has most of its mass near one---or, above the chosen $\delta$---when $\hat{\gamma}_r \in T(\hat{\gamma}_o;\boldsymbol{\alpha}_T)$. Analysts should strive to set $M$ so that $S^R/M$ satisfies this desideratum. However, analysts are unlikely to want or be allowed to submit the same query of $\mathbf{D}^*$ multiple times, e.g., to estimate a sampling distribution of $S^R$ at one or more values of $M$. Such repeated queries use up the privacy budget, which may be limited if the data holder enforces a total budget per analyst or an overall budget on analyses with $\mathbf{D}^*$.  Instead, we recommend that analysts approximate the sampling distribution of $S^R/M$ using the following simulation.  We suppose that the published study provides the estimated standard error,  \(\hat{\sigma}(\hat{\gamma}_o)\), and that the analyst has selected a tolerance region $U(\hat{\gamma}_o;\boldsymbol{\alpha})$.  The simulation uses only these quantities to avoid using additional privacy budget.

\begin{itemize}
\item
  Step 1: Construct a grid of plausible values for \(\gamma_r\) and potential values of \(M\). 
  \item Step 2: 
  For a given $(\gamma_r, M)$, simulate \(\hat{\gamma}_{r,l} \sim \mathcal{N} (\gamma_r,(n_0/ n)\hat{\sigma}^2(\hat{\gamma}_o))\) for \(l = 1, \dots, M\). Using these simulated $\hat{\gamma}_{r,l}$, compute \(S\) and sample \(\eta \sim Laplace(0,1/\epsilon)\), resulting in a draw of \(S^R = S+\eta\). Repeat this process independently a large number, say \(k=1000\), of times. The result is an approximate sampling distribution of $S^R/M$ for this \((\gamma_r, M)\).
\item
  Step 3: Repeat Step 2 for all combinations of $(\gamma_r, M)$.
\item Step 4:  Select an $M$ that offers a high probability of observing large values of $S^R/M$ for values of $\hat{\gamma} \in T(\hat{\gamma}_o, \boldsymbol{\alpha}_T)$ and small values of $S^R/M$ for values of $\hat{\gamma} \notin T(\hat{\gamma}_o, \boldsymbol{\alpha}_T)$.
\end{itemize}
As a visualization, for each $M$ under consideration, analysts can plot the median, 2.5\%, and 97.5\% quantiles of the distribution of $S^R/M$ at each value of \(\gamma_r\) in the grid in Step 1. The plot reveals the values of \(\gamma_r\) for which $S^R/M$ is likely to be near zero or one, i.e., the analysis unequivocally disfavors or favors stability. We illustrate the use of this simulation approach for choosing \(M\) and interpreting the results of AD stability analyses in Section \ref{illustration}.

We emphasize that this simulation is an approximation to the sampling distribution of $S^R/M$.  Stability analysts still should run the differentially private AD measure(s) on $\mathbf{D}^*$ as the ultimate analysis.

\subsubsection{Choosing $\delta$}

The MCMC sampler generates a set of $J$ posterior samples, $\{(r_{j},S_{j}): j=1, \dots, J\}$. Using these draws of $r$, the analyst can decide on a value $\delta$ of $r$ that represents sufficient evidence to conclude favorably for stability and compute the posterior probability, \(Pr(r \geq \delta | S^R) \approx \sum_{j=1}^J \mathbb{I}(r_j \geq \delta)/J\). When this probability is close to one, the analyst has evidence for stability ($r$ likely exceeds $\delta$), whereas when it is close to zero the analyst has evidence against stability ($r$ likely is less than $\delta$).

The choice of $\delta$ is specific to the  analyst's demands on evidence. Nonetheless, as a default, we suggest  \(\delta = 0.5\). With \(\delta = 0.5\), the analyst favors stability when the tolerance criterion is satisfied in the majority of partitions.  Additional theoretical support for using $\delta = 0.5$ is available in \cite{yangmsthesis}.

\subsection{Implementation of AM measures}\label{sec:workflowAM}

Generally speaking,  when choosing $M$ for the AM measures we confront trade offs like those in Section \ref{sec:workflowAD}. With larger \(M\), the average overlap \(\bar{\nu}\) is based on more partitions, which decreases the variance from the random partitioning and reduces the variance of the Laplace noise. However, the lengths of the CIs within the partitions increase because of the increased standard errors. As a result, for large \(M\), we can lose sensitivity to detect differences in $\beta$  and $\gamma$. Thus, stability analysts should seek an \(M\) that  controls the influence of the Laplace noise and the sampling variance of \(\bar{\nu}\) while preserving enough sensitivity to detect meaningful differences in $\beta$ and $\gamma$.

Before turning to the selection of $M$, it is worthwhile to consider the interpretation of $\bar{\nu}$.  First, let $\nu$ be the overlap measure for the CIs computed with all $N$ observations in $\mathbf{D}^*=\mathbf{D}$, that is, if there were no privacy concerns and no partitioning.  Critically, $\bar{\nu}$ is not necessarily an approximation of $\nu$. For example, suppose $|\beta - \gamma|>0$ but small, and $N$ is very large. In this case, it is likely that $\nu=0$. On the other hand, in this same setting, if one makes $M$ very large resulting in small $n$, in expectation $\bar{\nu}$ may not equal zero, as the inflated standard errors from small $n$ can cause the CIs to overlap.  Thus, analysts must interpret $\bar{\nu}$ as a distinct quantity from $\nu$.

When selecting $M$, one desideratum is that $\bar{\nu}^R$ is small when $\gamma$ and $\beta$ differ greatly and  large when they do not.  Thus, it is helpful to assess the sampling distribution of $\bar{\nu}^R$ for various differences in  $\gamma$ and $\beta$ at different values of $M$. Analysts can identify an $M$ that offers good differentiation in the values of $\bar{\nu}^R$ across practically meaningful differences in $\gamma$ and $\beta$.  With this goal in mind, we now present a simulation approach to facilitate such investigations without using any additional privacy budget.  Here, we presume that $\mathbf{D}=\mathbf{D}^*$ and that values of $\hat{\gamma}_o$ and $\hat{\sigma}^2(\hat{\gamma}_o)$ have been published.

We assume that the joint distribution of the estimated coefficients for \(X\) in any random partition of size $n$ can be approximated as  
\begin{equation}\label{eq:hats}
\begin{pmatrix}\hat{\gamma}_l \\ \hat{\beta}_l \end{pmatrix} \sim \mathcal{N}\left(\begin{pmatrix}\gamma \\ \beta \end{pmatrix}, \begin{pmatrix}\sigma^2(\hat{\gamma}_l) & \sigma(\hat{\gamma}_l,\hat{\beta}_l) \\ \sigma(\hat{\beta}_l,\hat{\gamma}_l) & \sigma^2(\hat{\beta}_l)  \end{pmatrix}\right).
\end{equation}
We set $\gamma=\hat{\gamma}_o$ and \(\sigma^2(\hat{\gamma}_l) =  (N/n)\hat{\sigma}^2(\hat{\gamma}_o)\). For  $\beta$ and $\sigma^2(\hat{\beta})$, the analyst posits plausible values for the percent difference in the coefficients, \(|\gamma - \beta|/|\gamma|\), as well as plausible ratios of their standard errors, \(\sigma(\hat{\gamma}_l)/\sigma(\hat{\beta}_l)\).  These can be used to solve for values of $\beta$ and $\sigma^2(\hat{\beta})$.  Unfortunately, the covariance \(\sigma(\hat{\gamma}_l, \hat{\beta}_l)\) is not  easily determined under $\epsilon$-DP, as directly using the estimated correlation between the $M$ pairs of $(\hat{\gamma}_l,\hat{\beta}_l)$ violates the privacy guarantee. As a simple default, analysts can set $\sigma(\hat{\gamma}_l,\hat{\beta}_l)=0$.  This tends to result in lower than actual simulated CI overlap than when $\sigma(\hat{\gamma}_l,\hat{\beta}_l)> 0$, which is expected since the correlation among the $(\hat{\gamma}_l,\hat{\beta}_l)$ is generally positive.

With this in mind, the simulation proceeds as follows.
\begin{itemize}
    \item Step 1:  The  analyst specifies a grid of plausible values of \(|\gamma - \beta|/|\gamma|\) and  \(\sigma(\hat{\gamma}_l)/\sigma(\hat{\beta}_l)\).
    \item Step 2:  For a given $(|\gamma-\beta|/|\gamma|, \sigma(\hat{\beta}_l)/\sigma(\hat{\gamma}_l))$, the analyst generates $M$ draws of $(\hat{\gamma}_l, \hat{\beta}_l)$ from \eqref{eq:hats}.  For each drawn  $(\hat{\gamma}_l, \hat{\beta}_l)$, the analyst computes the 95\% CI for each estimand marginally using the standard errors from \eqref{eq:hats} based on normal distributions.
    \item Step 3:  The analyst computes \(\nu_l\) for $l=1, \dots, M$, and subsequently \(\bar{\nu}\).  The analyst adds noise with the Laplace Mechanism to create  \(\bar{\nu}^R\). 
    \item Step 4:  The analyst repeats Step 2 and Step 3 a large number, say $k=1000$, of times.  This is the simulated sampling distribution of \(\bar{\nu}^R\) at  the given $(|\gamma-\beta|/|\gamma|, \sigma(\hat{\beta}_l)/\sigma(\hat{\gamma}_l)).$ 
    \item Step 5:  The analyst repeats Step 2 through 4 for all values of $(|\gamma-\beta|/|\gamma|, \sigma(\hat{\beta}_l)/\sigma(\hat{\gamma}_l))$ in the grid.
\end{itemize}
For different values of $M$, analysts can visualize the results over the grid points via a plot of the means of $\bar{\nu}^R$ over the $k$ simulations. The analyst can select a value of $M$ that offers adequate differentiation in the posterior means across different values of $\beta$.  We illustrate this process when applying the AM framework in Section \ref{illustration}, to which we now turn.

\section{Illustrative Analysis}\label{illustration}

We illustrate the stability measures using the 2019 American Community Survey data from the IPUMS USA database \citep{Steven2021ipums}. We discard individuals with any of the following characteristics: income less than \$1; unemployed; age outside the range 26 to 65 years; has difficulty in independent living; highest education less than one-year high school attendance; or, missing values in any of the features. After we recode some variables for analysis purposes, the resulting data comprise 1,175,526 individuals measured on the individual's total income in dollars, age in years, number of people in the family, sex (2 levels), marital status (2 levels), race (3 levels), health insurance status (4 levels), employment type (2 levels), veteran status (2 levels), and  highest education (less than college vs. at least one-year college). We subset and recode this way for simplicity of illustration rather than as a best practice for analyzing the IPUMS USA data. 

The analysis aims to study the relationship between income and highest education, i.e., some college or not, controlling for the other variables.  We presume results from a linear regression addressing this question already have been published.  We act like an analyst with access to the full 1,175,526 records under a confidentiality agreement who cannot release stability results unless they are differentially private. For all results, we use an $\epsilon=1$ per analysis.

\subsection{AD analysis}\label{illustration:AD}

Suppose the original researchers estimated a linear regression with log(income) on main effects of all the aforementioned covariates using all $n_o=1,175,526$ individuals.  They published the estimated coefficient of the indicator of some college attendance \(\hat{\gamma}_o = 0.459\) with estimated standard error \(\hat{\sigma}(\hat{\gamma}_o) = 1.7\times 10^{-3}\). Thus, $\hat{\gamma}$ suggests that, holding all the covariates constant, attending at least some college is associated with an approximately 0.459 higher log salary on average, which can be interpreted as a  $e^{0.459}-1=58\%$ higher median salary.

The regression model assumes that the true coefficient is the same for all subpopulations.  To probe this assumption, a researcher seeks to examine the effect of college attendance for the subgroup defined by  $N=557,397$ individuals in small families (no more than 2 members).  We suppose this researcher considers two stability queries. First, the researcher uses the adjusted tolerance region in \eqref{eq:U:AD:CI}, with  $(\hat{\gamma}_o, \hat{\sigma}(\hat{\gamma}_o))$ from the published study results and $\alpha = 3$.  Coefficients not in this range are roughly at least three standard errors away from the original estimate, indicating $\gamma_r$ may differ from $\gamma_o$ for this population.  Second, in the event that the first stability check fails, the researcher assesses the practical significance of any potential difference using a fixed tolerance region of \(U = [0.413, 0.504]\), i.e., within $\pm 10\%$ of $\hat{\gamma}_o$.

\begin{figure}
\centering
\epsfig{figure=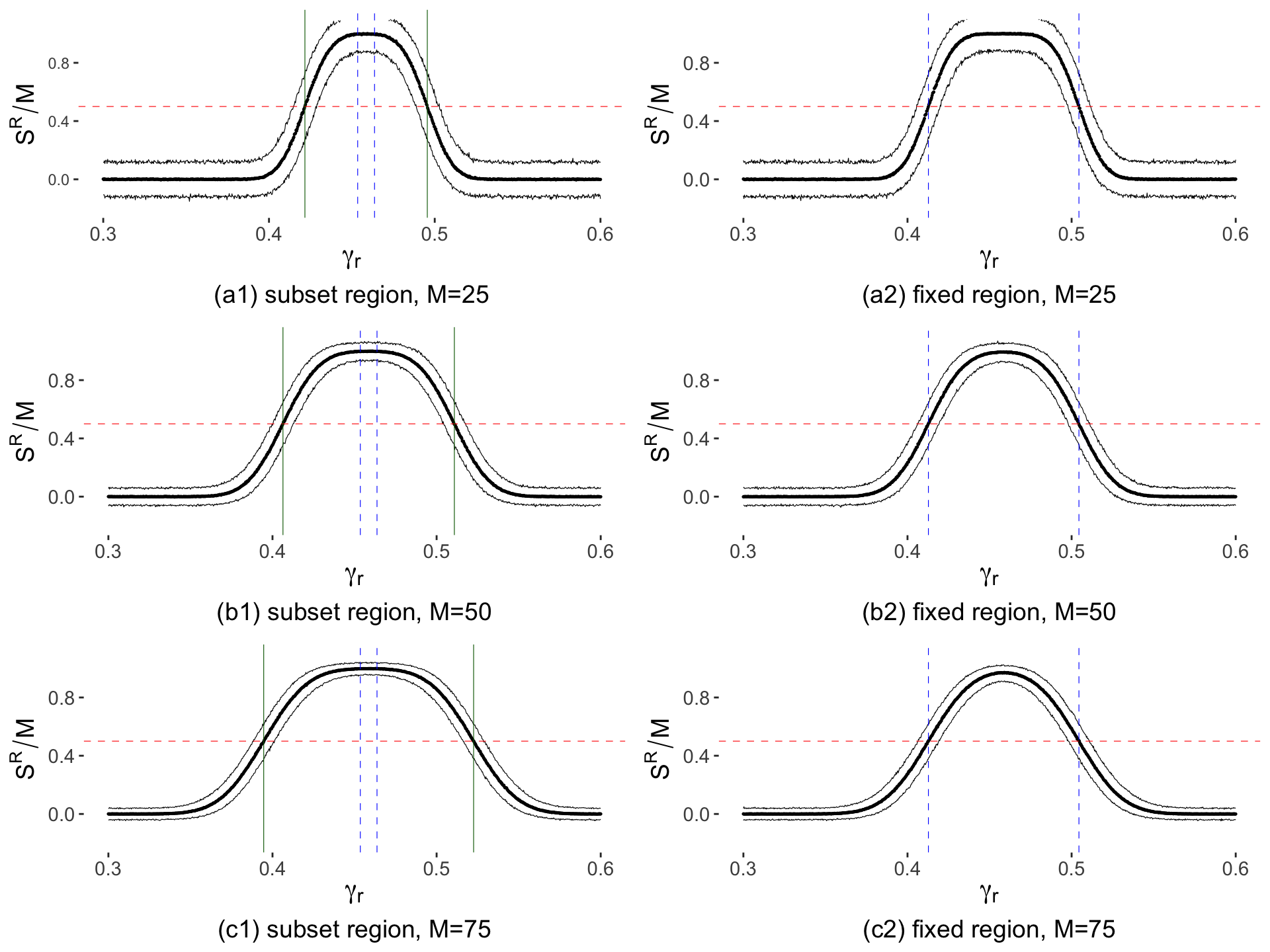, scale=0.24}
\caption{Simulated sampling distributions of \(S^R/M\) for the adjusted and fixed tolerance intervals for $M\in \{25,50,75\}$. Results are based on 1,000 randomly sampled values for each plausible $\gamma_r$.  We use a grid of values for $\gamma$ from 0.3 to 0.6 evaluated at increments of .0003.  Curves represent the 2.5\%, 50\%, and 97.5\% quantiles of the simulated values. Horizontal dashed lines indicate $\delta= 0.5$, and vertical dashed lines and solid lines indicate the limiting points of the corresponding $T(\hat{\gamma}_o, \alpha_T)$ and $U(\hat{\gamma}_o, \alpha)$, respectively.} \label{fig:SR:application}
\end{figure}

We select \(M\) for both the adjusted and fixed tolerance regions by referring to simulated sampling distributions of $S^R$, as described in Section \ref{sec:workflowAD}. Figure \ref{fig:SR:application} displays the sampling distributions for $M \in \{25, 50, 75\}$ based on \(\hat{\gamma}_o = 0.459\) and \(\hat{\sigma}(\hat{\gamma}_o) = 1.7\times 10^{-3}\).  We do not consider $M>75$ to avoid extra variability in $\hat{\gamma}_{r,l}$ due to relatively small $n$, and we do not consider $M< 25$ over concerns about the effect of the Laplace noise on the result. Based on the simulations in Figure \ref{fig:SR:application}, we select $M=25$ for the adjusted region for the following reasons.  First, the simulated values of $S^R/M$ are almost always well above $0.5$ when $\gamma_r$ is inside the tolerance region defined by \eqref{eq:T:AD:CI} with $\alpha_T=3$, i.e., $0.459 \pm 3 \times 1.7\times 10^{-3}$. Second, when $\gamma_r$ is outside this tolerance region, using $M=25$ offers the highest chance that $S^R/M<0.5$.  For the fixed tolerance region, we again select $M=25$.  Outside the tolerance region, the values of $S^R/M$ go to zero fastest when $M=25$.

\begin{figure}
\centering
\epsfig{figure=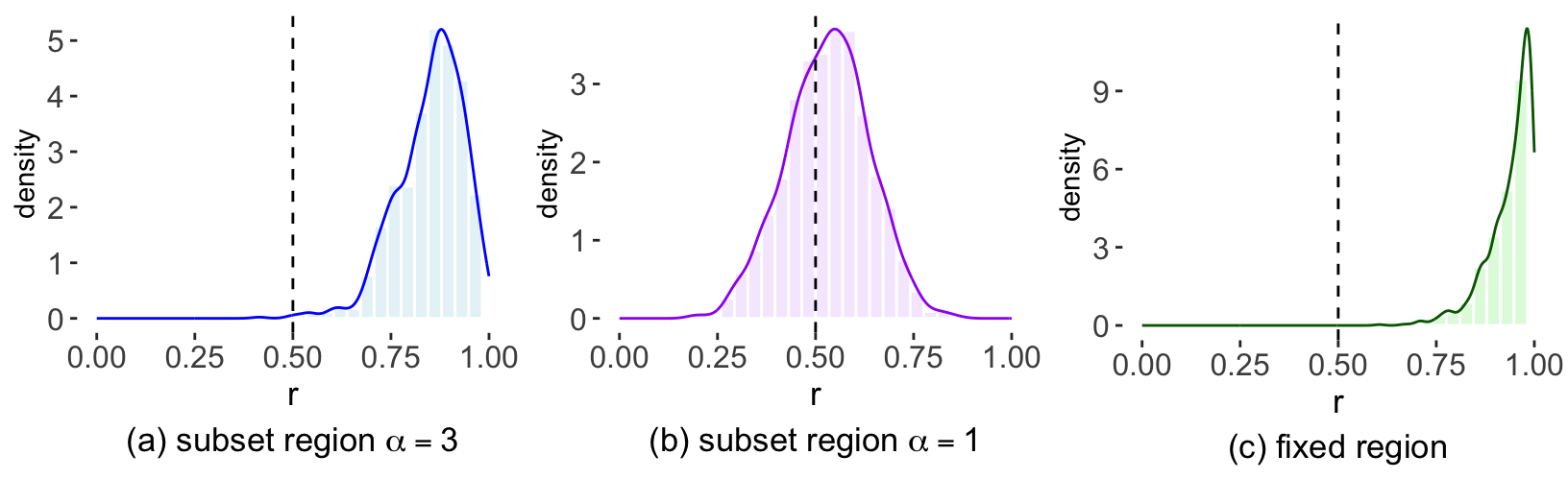, scale=0.27}
\caption{Posterior distribution of \(r\) for adjusted and fixed tolerance intervals for the subgroup of small families. Dashed lines indicate $\delta= 0.5$.} \label{fig:r:application}
\end{figure}

Applying the AD method with the adjusted tolerance region in \eqref{eq:U:AD:CI}, we obtain $S^R/M = 0.89$. We generate 1500 samples from the posterior distribution of $r|S^R$, discarding the first 500 as burn-in. Figure \ref{fig:r:application}(a) displays the posterior distribution of $r$.  Its posterior median is 0.86; the 95\% credible interval for $r$ is $(0.67,0.98)$; and, the $Pr(r>0.5|S^R) = 1$.  Evidently, the estimated $\hat{\gamma}_{r,l}$ for the subgroup of small families usually are inside the adjusted tolerance interval. This suggests that $\gamma_r$ is likely close to $\gamma_o$, and that the stability check passes.  The $\hat{\gamma}_o$ is a reasonable estimate of the coefficients for small families as well.

It is instructive to examine the results for a  researcher who demands closer estimates to claim stability. Suppose that instead the researcher uses the adjusted tolerance interval with $\alpha=1$. In this case, we have $S^R/M = 0.69$.  The 95\% credible interval for $r$ is $(0.31,0.73)$, and $Pr(r>0.5|S^R) = 0.62$. This is an ambiguous result: the evidence is not overwhelmingly in favor of or against the stability.  In this case, the researcher cannot feel confident that the stability criterion is satisfied.

Given the stability result and the interval length when $\alpha=3$, the researcher need not continue with the fixed tolerance interval check.  With  $\alpha=1$, however, they may want to assess whether differences in $\gamma_r$ and $\gamma_o$ are practically significant.  Thus, for completeness, we also  present the results for the fixed tolerance region of $U = [0.413, 0.504]$.  The value of $S^R/M = 0.99$.  Figure \ref{fig:r:application}(c) displays the posterior distribution of $r$ for this measure.  The posterior median of $r$ is 0.96; the 95\% credible interval is $(0.78, 1.00)$; and, the $P(r \geq 0.5 | S^R) = 1$. Thus, the researcher concludes that $\hat{\gamma}_r$ does not decrease or increase  more than 10\%; the difference is not practically significant according to this fixed tolerance interval.

We can use the simulated sampling distributions in  Figure \ref{fig:SR:application}  with $M=25$ to facilitate additional interpretations of the results. For the adjusted tolerance region, we see that $S^R/M$ is likely to exceed 0.5 for $0.43 < \gamma_r < 0.48$, and it is likely to dip below 0.5 for $\gamma_r<0.42$ or $\gamma_r>0.51$.  For the fixed tolerance region, we see that $S^R/M$ is likely to exceed 0.5 for $0.43 < \gamma_r < 0.50$, and it is likely to dip below 0.5 for $\gamma_r< 0.41$ or $\gamma_r> 0.51$.  The observed value of $S^R/M = 0.89$ for the adjusted tolerance region with $\alpha=3$ suggests that $\hat{\gamma}_r$ lies somewhere near 0.46; this is also supported by the large observed value of $S^R/M = 0.99$ for the fixed tolerance region.  In fact, restricting the linear regression to the small families, we have $\hat{\gamma}_r = 0.453$ with a standard error of $2.4\times 10^{-3}$.  Thus, the AD measures indeed lead us to reasonable conclusions about the stability analysis for this subgroup.

\subsection{AM analysis}\label{illustration:AM}

To illustrate an AM analysis, we now suppose the original analysis is conducted on the full data set, i.e., $\mathbf{D}$ comprises $N=1,175,526$ individuals.  The original researchers estimated a linear regression with main effects only, which is \(Model_0\), from which they published \(\hat{\gamma}_o = 0.459\) and \(\hat{\sigma}(\hat{\gamma}_o) = 1.7\times 10^{-3}\).  The stability researcher wants to fit an alternative model on  \(\mathbf{D}\) that, in addition to the predictors in \(Model_0\), includes interaction terms of age with the dummy variable for sex, age with the dummy variables for race, and age with the dummy variables for marital status.  We identified these interactions as potentially relevant based on exploratory analysis with $\mathbf{D}$.  Thus, $\beta$ is the coefficient for at least some college in the model that includes these added interaction effects.

Given $N$, we do not want $M$ to be so large that a small $n$ causes high variability in the CIs; see the supplementary material for simulations that demonstrate how small $n$ can make it difficult for $\bar{\nu}^R$ to show differences between $\gamma$ and $\beta$. Therefore, following the suggestions in Section \ref{sec:workflowAM}, we consider \(M \in \{25, 50\}\). These values are large enough to provide adequate gradation in $\bar{v}$, keep large $n$, and have reasonable global sensitivity for the sub-sample and aggregate mechanism. We follow the procedure in Section \ref{sec:workflowAM} and  simulate values of $\bar{\nu}^R$ across a grid of relative absolute differences in the coefficients and ratios of the standard errors ranging from 0.8 to 1.2.  For the absolute differences, because of the small standard errors and large sample size, we find that $\bar{\nu}$ is very small whenever the percent difference $|\beta-\gamma|/|\gamma|$ exceeds 15\%. Evidently, for this simulation design, the stability analysis nearly always can detect such very (and arguably unrealistically) large differences in the effects of college education on income. Thus, to aid the visualization, we only report results over  a grid between 0 and 0.15, using 300 equally spaced values of $|\beta-\gamma|/|\gamma|$,  i.e.,  increments of 0.0005.  The range of 0.8 to 1.2 for the standard deviation ratio reflects the belief that adding a few interaction terms to \(Model_0\) is unlikely to change the standard error dramatically. Within this range, we use 80 values at increments of 0.005.

\begin{figure}
\centering
\epsfig{figure=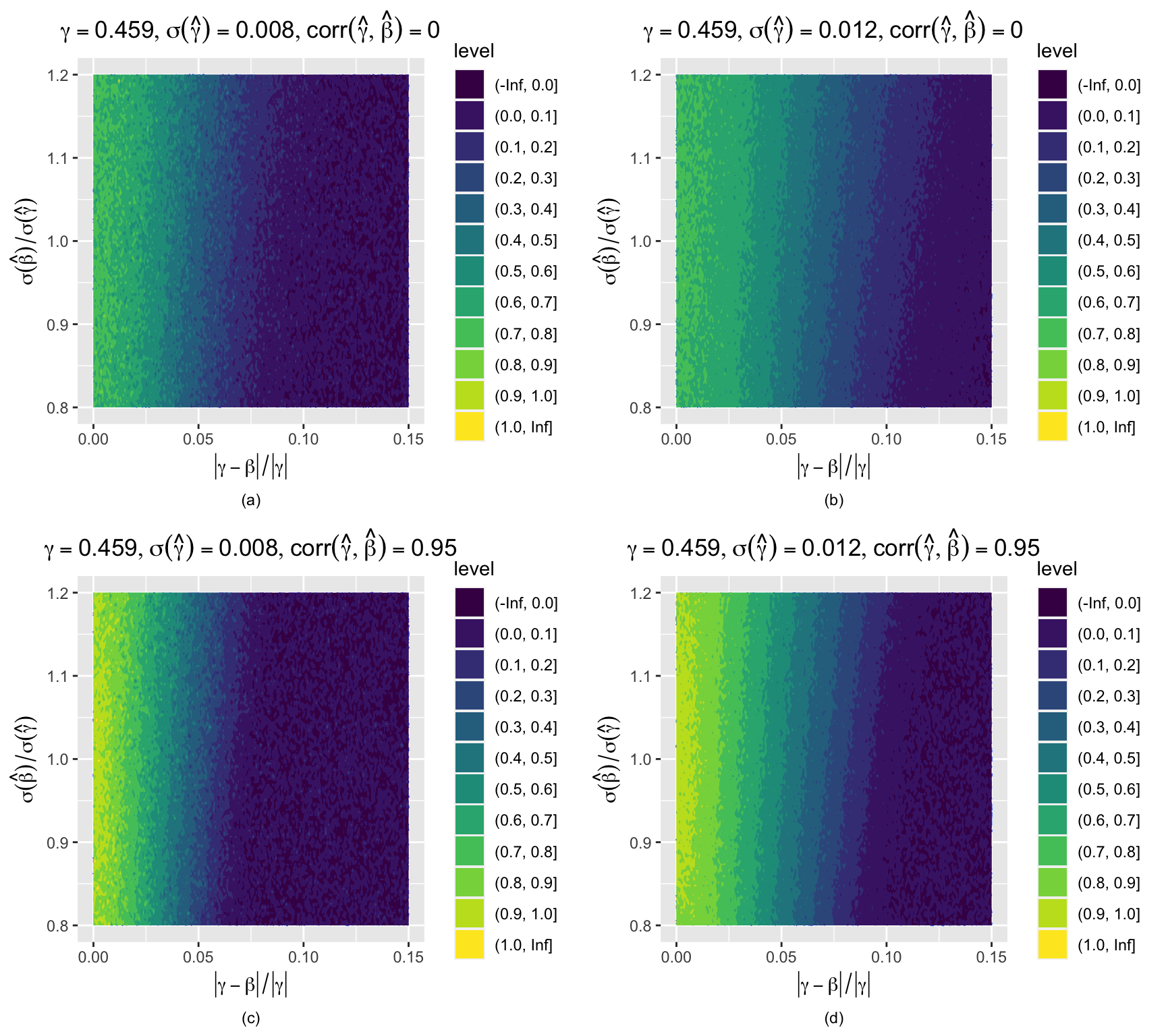, scale=0.265}
\caption{Simulated values of \(\bar{\nu}^R\) for the two models in the AM illustration.  Plots (a) and (c) use \(M=25\), and plots (b) and (d) use \(M=50\).  Plots (a) and (b) presume a correlation between $\hat{\gamma}_l$ and $\beta_l$ equal to zero, and plots (c) and (d) presume a correlation equal to 0.95. $\sigma(\hat{\gamma})$ is an approximate adjusted standard error based on the subset sample size.  Results are derived from the averages of 500 simulated  \(\bar{\nu}^R\) values at each point on the grid.} \label{fig:AM:app}
\end{figure}

Figure \ref{fig:AM:app} displays the simulated average values of $\bar{\nu}^R$ over the grid under two scenarios for each \(M\). One presumes that the correlation between \(\hat{\gamma}_l\) and \(\hat{\beta}_l\) equals zero, and the other presumes that this correlation equals 0.95. With either correlation, $M=25$ and $M=50$ both offer adequate differentiation in $\bar{\nu}^R$. Considering that $M=50$ produces a Laplace distribution with smaller variance and offers a finer resolution, we select $M=50$ for the AM stability analysis.

Figure \ref{fig:AM:app} also illustrates some properties of the AM measure.  First, it verifies that the AM measure offers more nuanced results for positive correlations than zero correlation. Second, it shows that the measure does not differentiate some values of $|\gamma-\beta|/|\gamma|$ but does others. For example, in Figure \ref{fig:AM:app}, we expect the AM measure with $M=50$ to take on similar, small values for any relative differences greater than 10\%.  In contrast, we expect the measure to be relatively sensitive to changes in $|\gamma-\beta|/|\gamma|$ for  differences of 5\% and lower. Of course, these conclusions are specific to the published $(\hat{\gamma}_o,\hat{\sigma}(\hat{\gamma}_o))$. Finally, analysts can utilize  plots like Figure \ref{fig:AM:app} to facilitate interpretations of the AM measure in their specific setting.  To illustrate, consider a value of $\bar{\nu}^R$ with $M=50$ around 0.8.  In Figure \ref{fig:AM:app}, this would suggest that  $|\beta - \gamma|/|\gamma|$ is in the neighborhood of 2.5\%.  On the other hand, a value of $\bar{\nu}^R$ with $M=50$ around 0.1 would suggest that  $|\beta - \gamma|/|\gamma|$ is at least 10\%.

Having selected $M=50$, we implement the differentially private algorithm once and obtain a \(\bar{\nu}^R \approx 0.94\).  We then sample 1000 posterior draws of \(\bar{\nu}\) with the default prior \(\psi_0 = Beta(1,1)\). The posterior median of $\bar{\nu}$ is 0.94 with a 95\%  credible interval of \([0.83, 0.99]\).  These results suggest that inferences for $\beta$ and $\gamma$ are not meaningfully different for these two models.  Indeed, this is actually the case:  for \(Model_1\)  $\hat{\beta}= 0.456$ with a standard error of $1.7\times 10^{-3}$.  The relative difference in these estimates is less than 1\%, which is roughly the difference Figure \ref{fig:AM:app} tells us to expect for \(\bar{\nu}^R \approx 0.94\).

\subsection{Repeated sampling evaluations}

As a check on the results, we examine the repeated sampling performances of the AD and AM measures.  For the AD measures, we repeat the process of generating $S^R$ 200 times and compute the posterior distribution of $r$ in each simulation run. This is not feasible under differential privacy, at least for a reasonable privacy budget over all queries, but it does allow us to assess if the results from Section \ref{illustration:AD} are representative.  For the measure with the adjusted tolerance region with $\alpha=3$,  95\% of the 200 values of $Pr(r\geq 0.5 | S^R)$ are within $(0.998, 1)$; with $\alpha=1$, 95\% of the 200 values of $Pr(r\geq 0.5 | S^R)$ are within $(0.46, 1)$.  For the measure with the fixed tolerance region, 95\% of all 200 values of  $Pr(r\geq 0.5 | S^R)$ are also within $(0.998,1)$. Thus, our one-time results for both the adjusted and fixed regions are not atypical. 

For the AM measure, we repeatedly generate 200 values of $\bar{\nu}^R$ and for each find the posterior distribution of $\bar{\nu}$. Approximately 95\% of the 200 values  of $\bar{\nu}^R$ are between $(0.89, 1.00)$. Using the values of $\bar{\nu}^R$  at these end points, the 95\% central credible intervals for $r$ are $(0.77,0.98)$ and $(0.85,1.00)$, respectively. Thus, our one-time result is not atypical.

\subsection{Comparing AD with a direct application of the Laplace mechanism}\label{sec:advslm}

Having illustrated the AD and AM stability measures, we return to a question posed at the beginning of this article: should analysts release  differentially private outputs directly rather than use an AD or AM stability measure?   In this section, we present  numerical evidence that the AD stability measures can be more effective than reporting a noisy result from the Laplace mechanism directly.

To so do, we use as the original (published) estimate from $\mathbf{D}$ the average income among the 1,175,526 individuals, which is  $\bar{y}=69834$. We suppose that the  researcher examines whether the average income varies across some subgroups and sets $69834 \pm 2500$ as the tolerance region. When the mean income $\bar{y}_s$ for a subgroup $s$ is outside this region, the researcher claims that the subgroup mean is sufficiently different from that of $\bar{y}$. Here, we consider the small families subgroup and also a subgroup comprising all the women in the data (566,599 individuals).

We implement the AD method with $\epsilon = 1$ and a fixed tolerance interval for $M \in \{25, 50\}$,  including only the intercept in the regression model.  The estimated intercept under least squares is the sample mean of the input data.  We also apply a Laplace Mechanism directly to each $\bar{y}_s$. To compute its global sensitivity, we assume that personal income is positive and bounded by \$1 billion.  Thus, in the Laplace Mechanism with $\epsilon=1$, we generate the perturbations from $\eta \sim Laplace(0, 1000000000 /n_s)$, where  $n_s$ is the sample size for  subgroup $s$. We repeat this process 1000 times.

For the female subgroup, $\bar{y}_s = 55222$. With such a large difference in means ($\bar{y}-\bar{y}_s = 14612$), both methods return the correct answer that $\bar{y}_s$ is outside the tolerance region. For the AD stability measure, the largest $Pr(r\geq 0.5 | S^R)$ over 1000 replications is only 0.02 when $M=25$ and always equals 0 when $M=50$. For the direct Laplace Mechanism, the noisy mean is outside the tolerance interval in 999 of the 1000 repetitions.

However, this is not the case for the small families subgroup.  For this $s$, we have $\bar{y}_s=68643$, which is a difference from $\bar{y}$ of 1191.  The AD stability measure generally returns the correct answer that the subgroup mean is within the tolerance region.  The smallest $Pr(r\geq 0.5 | S^R)$  across the 1000 repetitions is 0.96 when $M=25$ and is 1 when $M=50$.   The Laplace Mechanism, on the other hand, results in a noisy estimate outside the tolerance interval in 307 of the 1000 trials.  Evidently, the Laplace noise is large enough to obscure the difference of \$1191 a sizeable fraction of times.

In this stability analysis, the AD measure appears to be more reliable than a direct application of the Laplace Mechanism.  Of course, this comparison depends on the features of the application at hand including, for example, the global sensitivity, sample size, tolerance interval, and $\epsilon$.  For quantities that  differentially private algorithms can  estimate directly, analysts can apply  these algorithms to the data generated in the  simulation studies, e.g., in Step 2 of the simulation in Section \ref{sec:AD:M}, and assess whether a direct application or AD measure is more likely to give an accurate conclusion about the stability analysis.

\section{Concluding Remarks}\label{conclusion}

The AD and AM measures allow analysts to publish results of stability checks and satisfy differential privacy.  The usefulness of the measures naturally depends on $\epsilon$, $N$, and $M$. These interact in complex ways, making simulation studies like those used to select $M$ a useful way to understand the properties of the methods for particular settings, as we illustrated in Section \ref{illustration}. However, we can make some general statements.  With higher privacy demands, i.e., smaller values of $\epsilon$, we expect the measures to have decreased ability to reveal stability failures.  With larger $N$, we expect the estimates of the coefficients in the partitions to be more precise (for a given $M$), which in turn allows for more reliable assessments of stability.  Finally, with increased $M$ (for a given $n$), we expect to reduce the effect of the Laplace noise on the interpretability of the measures.  These trends are illustrated for the AD measures in the supplementary material. 

Simulations like those in Figure \ref{fig:SR:application} and Figure \ref{fig:AM:app} can allow analysts to assess whether a potential stability analysis is sufficiently sensitive and hence worth spending privacy budget on. For example, to assess if a   particular $n$ is large enough to offer meaningful stability analyses, analysts can examine results of simulation studies like those described in Section \ref{sec:workflowAD} and Section \ref{sec:workflowAM} with different $M$ (and hence $n$).  When the values of $S^R/M$ or $\bar{\nu}^R$  have adequate variation over the range of parameter values deemed of interest (for at least one $M$), analysts can expect interpretable results from the stability measures.  When this is not the case, the stability analysis is unlikely to be informative, e.g., small values of $n$ are likely to lead to inability to detect differences.

These studies also can inform the choice of parameters in the stability measure beyond $M$.  For example, in the AD measure with the adjusted tolerance region, if the distribution of $S^R/M$ when $\alpha=3$ is flat and near 1 for values of $\gamma_r$ deemed meaningfully  different than $\gamma_o$, the analyst may decide to tighten the requirement for stability and set $\alpha = 1$.  The results also suggest that it may be possible to approximate the distribution of $\hat{\gamma}_r \mid S^R$ or  the $Pr(\hat{\gamma}_r \in T(\hat{\gamma}_o;\boldsymbol{\alpha}_T) \mid S^R$) from the simulated distributions. Developing and assessing the quality of such approximations is a worthy topic for future research.  

When stability measures indicate that the alternative and original analyses are sufficiently similar, analysts may be satisfied to report summaries of the posterior distribution for $r$ and be done.  When this is not the case, however, analysts must decide what, if any, additional steps to take. Translating stability results to percent differences, as we illustrated in Section \ref{illustration}, provides one way to give a rough sense of the differences without using additional privacy budget.

In our illustrative analyses, we did not enforce a total privacy budget.  This is a policy decision by the data holders.  A naive accounting of the privacy budget loss adds the $\epsilon$ used per each query (and any permissible queries of the confidential data). Future research could develop more refined measures of privacy loss, so as to facilitate repeated stability queries under adherence to a total privacy budget.

We designed the AD measures for settings where  $\mathbf{D}^*$ and  $\mathbf{D}$ comprise disjoint sets of individuals or where $\mathbf{D}^* \subset \mathbf{D}$. In some settings, the stability analyst may want to add individuals' data to $\mathbf{D}$, so that $\mathbf{D} \subset \mathbf{D}^*$.  While one can apply our AD measures, it is worth noting several features of this approach. Suppose we use partitions of $\mathbf{D}^*$ and estimate $\hat{\gamma}_o$ from $\mathbf{D}$, as in Section \ref{sec:AD}. Further, suppose the data added to create $\mathbf{D}^*$ from $\mathbf{D}$ include an observation, say $(y_k, {x}_{k1}, \dots, x_{kp})$, that has very high influence when included in the estimation; that is, $\hat{\gamma}_r$ and $\hat{\gamma}_o$ are very different. This $(y_k, {x}_{k1}, \dots, x_{kp})$ can be in only one of the partitions and hence affect only that single $\hat{\gamma}_{r,l}$. The AD measure easily could return $W_l=1$ for most $l$, even if this point completely changes $\hat{\gamma}_o$ when included in the analysis.  One alternative is to use partitions of the records in $\mathbf{D}$ and compute $\hat{\gamma}_o$ from $\mathbf{D}^*$, which has the additional records. In this case, most $\hat{\gamma}_{r,l}$ across the $M$ partitions could be noticeably different than $\hat{\gamma}_o$, which would lead the analyst to the appropriate conclusion. However, since this method computes $\hat{\gamma}_o$ anew rather than use an existing, published value, changing one record in $\mathbf{D}^*$ can affect $W_l$ for each of the $M$ partitions; thus, the algorithm may not offer the reductions in variance from the subsample and aggregate mechanism. Developing stability measures for situations where $\mathbf{D} \subset \mathbf{D}^*$ is a topic for future research.

\newpage

\bigskip

\section*{Acknowledgements}
This work is supported by a grant from the National Science Foundation (SES-2214756).

\section*{Disclosure Statement}
The authors report there are no competing interests to declare.

\bibliographystyle{natbib}
\bibliography{0_ref.bib}

\end{document}